\documentclass{jcgt}

\setciteauthor{Annika Oehri, Philipp Herholz and Olga Sorkine-Hornung}
\setcitetitle{Smooth ARAP}

\accepted{2025-01-18}{2025-04-21}{2025-06-06}{Eric Lengyel}{14}{1}{1}{1}{2025}
\seturl{https://jcgt.org/published/0014/01/10/}

\usepackage{xcolor}
\usepackage{overpic}
\usepackage{booktabs}
\usepackage{pgfplots}
\usepackage{dsfont}
\usepackage{amsmath}
\usepackage{soul}
\usepackage{units}

\newcommand{\M}{\mathcal{M}}
\newcommand{\N}{\mathcal{N}}
\newcommand{\E}{\mathcal{E}}
\newcommand{\V}{\mathbf{V}}
\newcommand{\Vset}{\mathcal{V}}
\newcommand{\T}{\mathcal{T}}
\newcommand{\R}{\mathbf{R}}
\newcommand{\smolrot}{\mathbf{R}}
\newcommand{\svdu}{\mathbf{U}}
\newcommand{\svdv}{\mathbf{W}}
\newcommand{\svds}{\mathbf{S}}
\newcommand{\svdsig}{\boldsymbol{\Sigma}}

\newcommand{\Lap}{\textbf{L}}
\newcommand{\Mass}{\textbf{M}}
\newcommand{\Reals}{\mathds{R}}
\newcommand{\lapvec}{\boldsymbol\ell}

\DeclareMathOperator*{\argmin}{argmin}

\newcommand{\figref}[1]{Fig.~\ref{#1}}
\newcommand{\equref}[1]{Eq.~\eqref{#1}}

\newcommand\norm[1]{\left\lVert#1\right\rVert}
\newcommand{\tran}{\raisebox{1.05ex}{$\scriptscriptstyle \mathsf{T}$}}

\begin{document}

\title{Higher Order Continuity for Smooth As-Rigid-As-Possible Shape Modeling}

\author{Annika Oehri \and Philipp Herholz \and Olga Sorkine-Hornung}

\teaser{
    \begin{overpic}[trim=0cm 0.0cm 0cm 0cm,clip,width=0.9\linewidth,grid=false]{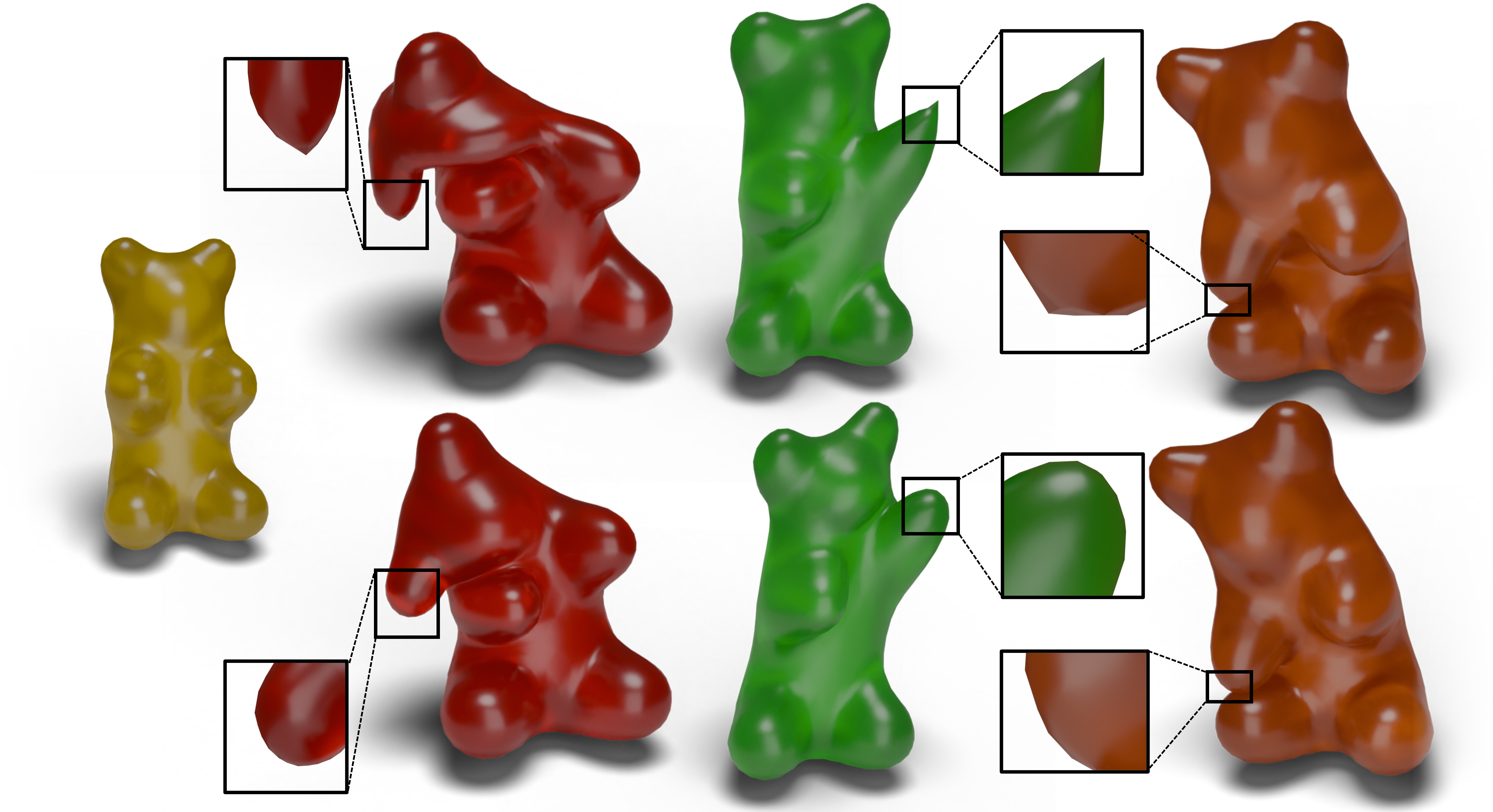}
    \put(3,44){\footnotesize standard}
    \put(3,41){\footnotesize ARAP}
    \put(3,11){\footnotesize smooth}
    \put(3,8){\footnotesize ARAP}
    \end{overpic}
  \caption{Deformations of a gummy bear, the original mesh depicted in yellow. The standard ARAP method {\protect \cite{arap}} can yield spikes, as visible in the highlighted regions. In contrast, our method produces smooth deformations. 
  }
  \label{fig:teaser}
}

\maketitle

\begin{abstract}

\small
We propose a modification of the As-Rigid-As-Possible (ARAP) mesh deformation energy with higher order smoothness, which overcomes a prominent limitation of the original ARAP formulation: spikes and lack of continuity at the manipulation handles. Our method avoids spikes even when using single-point positional constraints. Since no explicit rotations have to be specified, the user interaction can be realized through a simple click-and-drag interface, where points on the mesh can be selected and moved around while the rest of the mesh surface automatically deforms accordingly. 
Our method preserves the benefits of ARAP deformations: it is easy to implement and thus useful for practical applications, while its efficiency makes it usable in real-time, interactive scenarios on detailed models.

\end{abstract}

\section{Introduction}
\label{sec:introduction}
Shape deformation using user-defined, click-and-drag position constraints is a versatile method with numerous applications in industrial
and artistic design~\cite{DeformationTutorial:2009}. A prominent application is 3D shape modeling and animation for lay users, exemplified by the Monster Mash system \cite{MonsterMash:2020}. Monster Mash allows users to draw a 2D sketch of a shape, automatically inflates it into a 3D shape, and then enables users to interactively deform this mesh by simply clicking on surface points and moving them around, using As-Rigid-As-Possible (ARAP) shape deformation as the backbone \cite{arap}. 

\begin{figure}
    \centering
    \begin{overpic}[trim=0cm 0cm 0.5cm 0cm,clip,width=0.88\linewidth,grid=false]{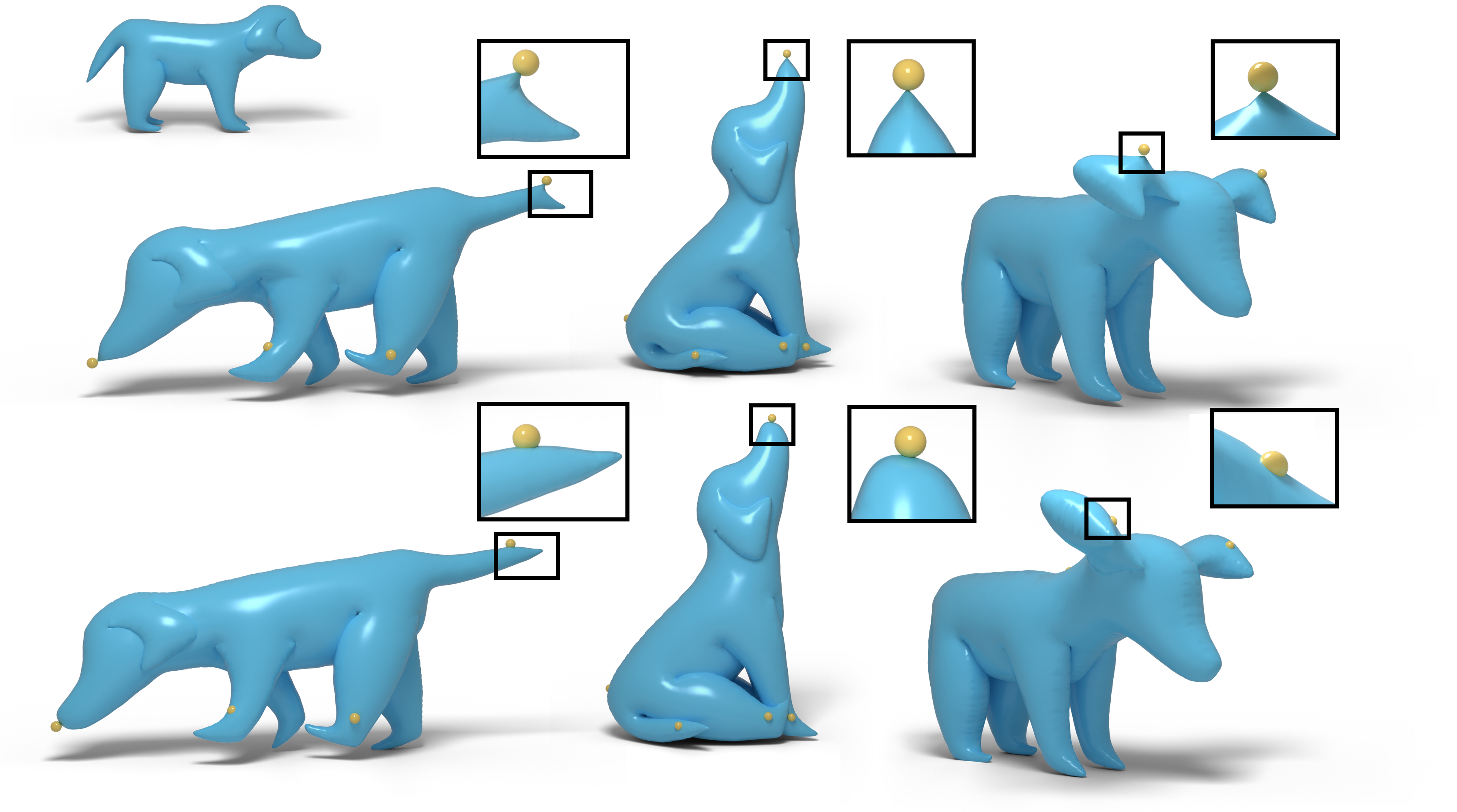}
    \put(6.5,44){\footnotesize original mesh}
    \put(-5,37){\footnotesize standard}
    \put(-5,34){\footnotesize ARAP}
    \put(-5,31){\footnotesize \shortcite{arap}}
    \put(-5,12){\footnotesize smooth}
    \put(-5,9){\footnotesize ARAP}
    \end{overpic}
  \caption{Interactive deformation of a dog model from the Monster Mash system {\protect \cite{MonsterMash:2020}} via single-point handles, indicated by yellow spheres. Some marked regions are shown in close-ups to better visualize the spikes produced by ARAP \protect\cite{arap} and the absence of spikes in our smooth formulation. }
  \label{fig:motivation_monster_mash}
\end{figure}

\begin{figure}
    \centering
    \begin{overpic}[trim=0cm 14cm 0cm 0cm,clip,width=1\linewidth,grid=false]{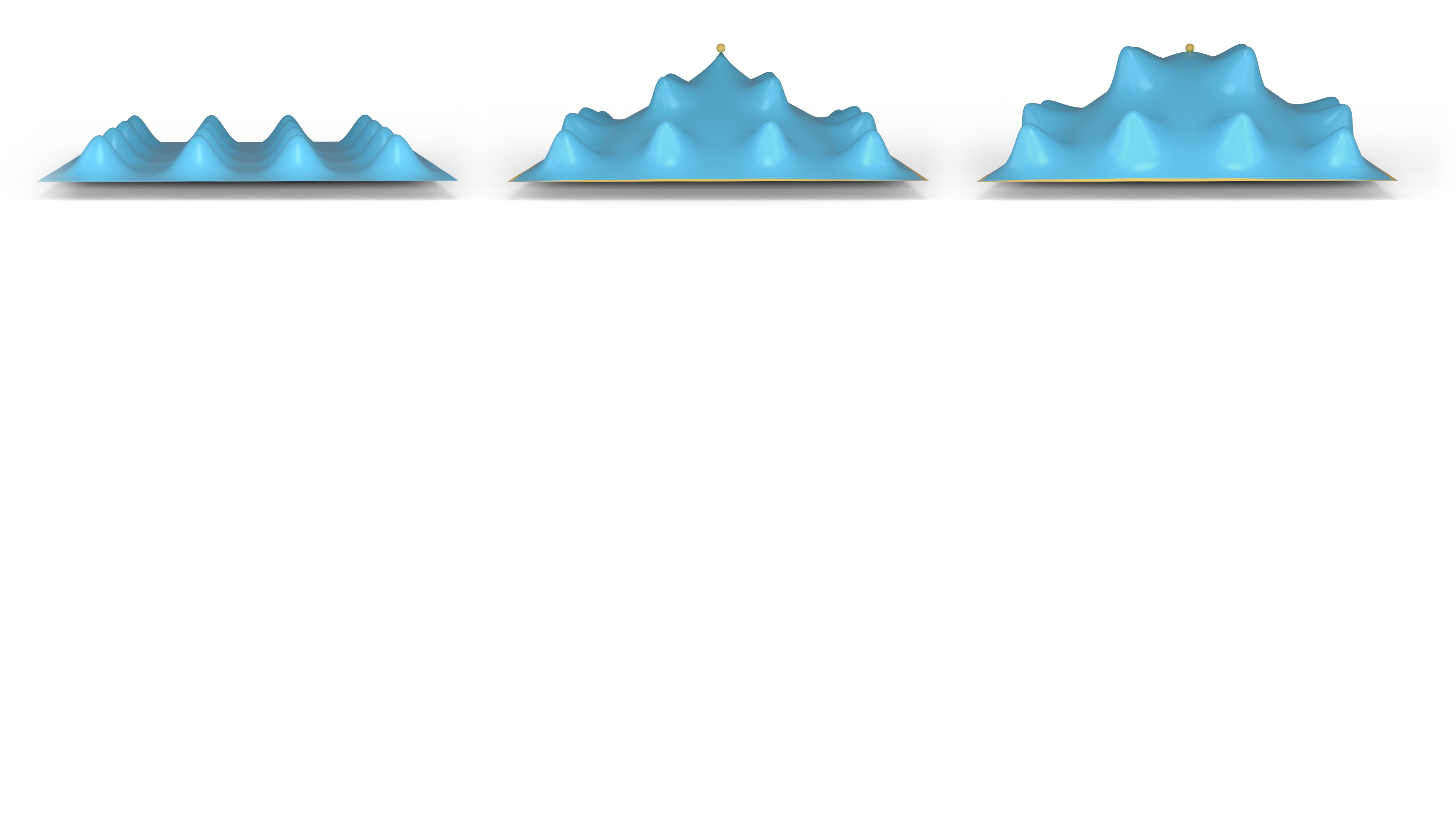}
    \put(10.5,14){\footnotesize original mesh}
    \put(38,14){\footnotesize standard ARAP \shortcite{arap}}
    \put(74,14){\footnotesize smooth ARAP}
    \end{overpic}
  \caption{Interactive deformation of a bumpy plane by dragging a point handle in the center of the mesh while fixing the boundary vertices. 
  }
  \label{fig:motivation_knubbel}
\end{figure}

ARAP \shortcite{arap} is a mesh deformation method that preserves the shape's overall appearance and geometric details by minimizing non-rigid local transformations, building on the fact that our understanding of shape is usually independent of its orientation and translation. 
The method is very efficient due to its formulation of the objective function, which, albeit highly nonlinear, can be quickly minimized via local-global optimization using sparse linear solvers. Additionally, a slight perturbation of the positional constraints usually causes only a small change to the resulting surface, making it especially useful for interactive applications such as Monster Mash. However, when deforming a mesh using the classical ARAP approach, spikes can appear, as depicted in Figures \ref{fig:teaser}, \ref{fig:motivation_monster_mash} and \ref{fig:motivation_knubbel}. This lack of smoothness at the constraints is particularly visible with fine mesh resolution and when using single-point handles. A smoother deformation is often desired and expected, where no newly introduced sharp features distort the shape.

We propose a modification of the original ARAP method that yields smoother results by raising the order of the objective function while staying within the efficient and easy-to-implement local-global optimization framework. Our method enables effective use of single-point handles, which greatly simplify the interaction metaphor of the deformation without introducing spikes. An added benefit of employing point constraints is the ability to perform fast updates of the underlying system matrix factorization, affording fast adding and removing of point handles and making the deformation overall more fluid and responsive to user interactions.

\subsection{Related work}
\label{sec:relatedwork}

As-Rigid-As-Possible shape deformation \cite{arap} is achieved by minimizing an objective function termed \emph{ARAP energy} that encourages local transformations of the surface to be as close to rigid as possible, while satisfying the positional constraints imposed by the manipulation handles. To minimize this nonlinear energy, a local-global optimization approach is adopted. In the local step, the optimal rigid transformation of each local neighborhood on the mesh is computed by solving the Procrustes problem. In the global step, the previously found rotations are kept fixed, and the vertex positions are updated by solving a sparse linear system. This process is iterated until convergence. The method is very robust, it is efficient thanks to precomputation and reuse of the sparse factorization of the system matrix and achieves compelling results through its automatic preservation of the relative orientation of local shape details. 
However, the deformed surface tends to lack tangent continuity at the positional constraints, and especially when single-point handles are used, spikes appear at the point constraints (see Figures \ref{fig:motivation_knubbel} and \ref{fig:motivation_monster_mash}). This phenomenon stems from the Poisson equation solved in the global step of ARAP, which cannot afford simultaneous positional and derivative constraints.  

Smoothness is commonly achieved by applying regularization. While many approaches tackle the problem in an energy-independent manner, it can be beneficial to consider the specific objective energy at hand. The method of Smoothed Quadratic Energies on Meshes \cite{smoothed_energy} offers a way of enforcing lower variation of the energy function that is by construction problem-specific. The proposal is to add a term that penalizes the total squared variation of the local energy, favoring minima where the energy is distributed more evenly across the whole mesh. This approach can be applied to different types of energies and application settings. Applied to ARAP, it yields a similar higher-order energy to ours, but with a different smoothing term, which uses the original ARAP neighborhood $\N$ instead of Laplacian vectors \eqref{eq:lap_vec} as we do. Martinez and colleagues ~\shortcite{smoothed_energy} focus mainly on 2D examples, while we demonstrate that our smooth ARAP performs well in complex and interactive 3D shape deformations. 

\begin{figure}
    \centering
    \begin{overpic}[trim=0cm 2cm 0cm 3.3cm,clip,width=\linewidth]{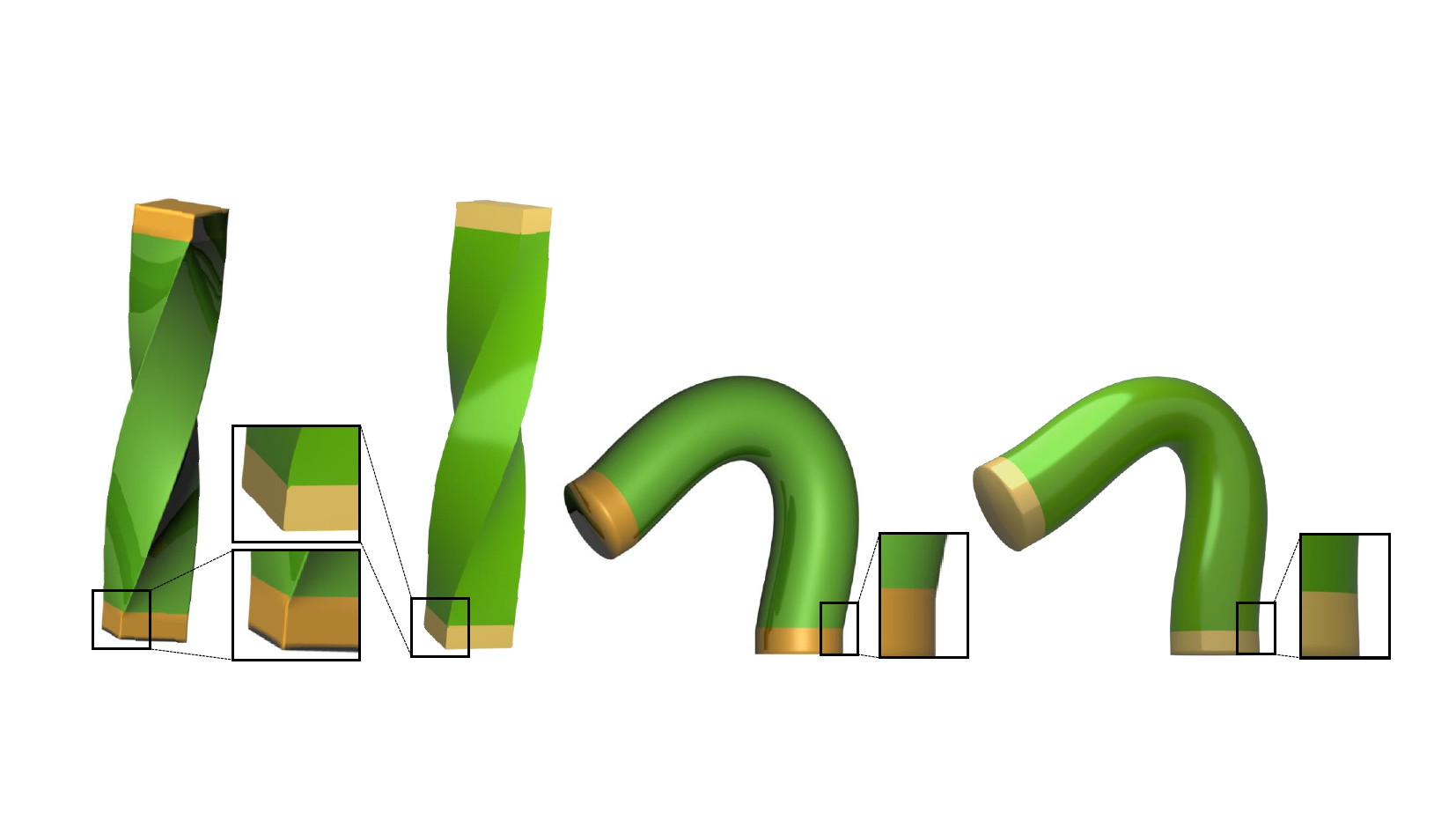}
        \put(5,0){\footnotesize SR-ARAP}
        \put(30.5,0){\footnotesize ours}
        \put(49,0){\footnotesize SR-ARAP}
        \put(81,0){\footnotesize ours}
    \end{overpic}
    \caption{Comparison between SR-ARAP and our smooth ARAP on the benchmark survey examples \protect\cite{DeformationSurvey:2008}. The SR-ARAP results are taken from Fig. 1 and 9 from the paper \protect\cite{Levi_smoothed} and our results are shown in a similar view. It can be seen that our global approach produces smoother transitions to the handles, while the locally smoothed rotations still have visible kinks, as highlighted.}
    \label{fig:sr_comp}
\end{figure}

In the work of Levi and Gotsman \shortcite{Levi_smoothed}, another form of smoothness regularization is proposed that specifically targets the surface deformation of ARAP, termed \emph{SR-ARAP energy}. 
The smoothness of rotations is achieved by adding a term to the surface energy that penalizes the difference between the rotations of neighboring edge sets, to encourage neighborhoods to transform as a unit. 
The local step of ARAP is adapted such that the best rotation is found not only based on the deformed edge positions but also on the previously computed neighborhood rotations.  This approach smooths rotations only locally, and the overall system remains Poisson-based, meaning positions from the global step may not reflect a smooth deformation, even with smooth rotations, as shown in \figref{fig:sr_comp}. Notably, they do not show results with single-point handles. In our method, we address the smoothness issue in the \emph{global} step, by raising the order of the PDE and employing the bi-Laplace operator, resulting in smooth deformation results even around point handles.

\section{Energy definition and method}

\paragraph{Problem statement and notation.}
Let $\M = (\Vset, \T)$ be a given manifold, orientable triangle mesh with $n$ vertices, where $\Vset$ is the vertex set with corresponding vertex positions $\V \in \Reals^{n\times3}$ and $\T$ is the face set, where we define the set of all triangles containing vertex $v$ to be $\T_v = \{t \in \mathcal T | v \in t\}$. Our goal is to determine the new deformed configuration $\M' = (\Vset', \T)$ satisfying some positional constraints, namely that the \emph{handle vertices} are at the specified positions. At the same time, we want to minimize an energy to preserve the shape locally while avoiding spikes. 

Let $A_v$ denote the Voronoi area of the vertex $v \in \Vset$ and let all such areas be accumulated on the diagonal of the lumped mass matrix $\Mass \in \Reals^{n\times n}$. We use a half-edge data structure for the mesh and denote by $\E \subset \Vset ^2$ the set of \emph{half-edges} (or \emph{directed edges}). Each half-edge belongs to a triangle, and we adopt the standard convention that the half-edges in a triangle are oriented counterclockwise. 
We use $e=(u,v)$ to refer to the half-edge $e$ going from vertex $u$ to $v$, and $\mathbf{e}=\mathbf{v}-\mathbf{u}$ to denote the corresponding vector in $\Reals^3$, which makes use of the vector representation $\mathbf{v}$ of vertex $v$.
Additionally, we distinguish between the original and deformed configurations by marking the deformed configuration with a prime. For example, $\mathbf{e},\mathbf{e}' \in \E$ both refer to the same half-edge $e$ in terms of mesh connectivity, but $\mathbf{e}'$ uses the updated vertex positions $\V'$. 

\begin{figure}
    \centering
    \begin{overpic}[trim=0cm 0cm 0cm 0cm,clip,width=0.5\linewidth,grid=false]{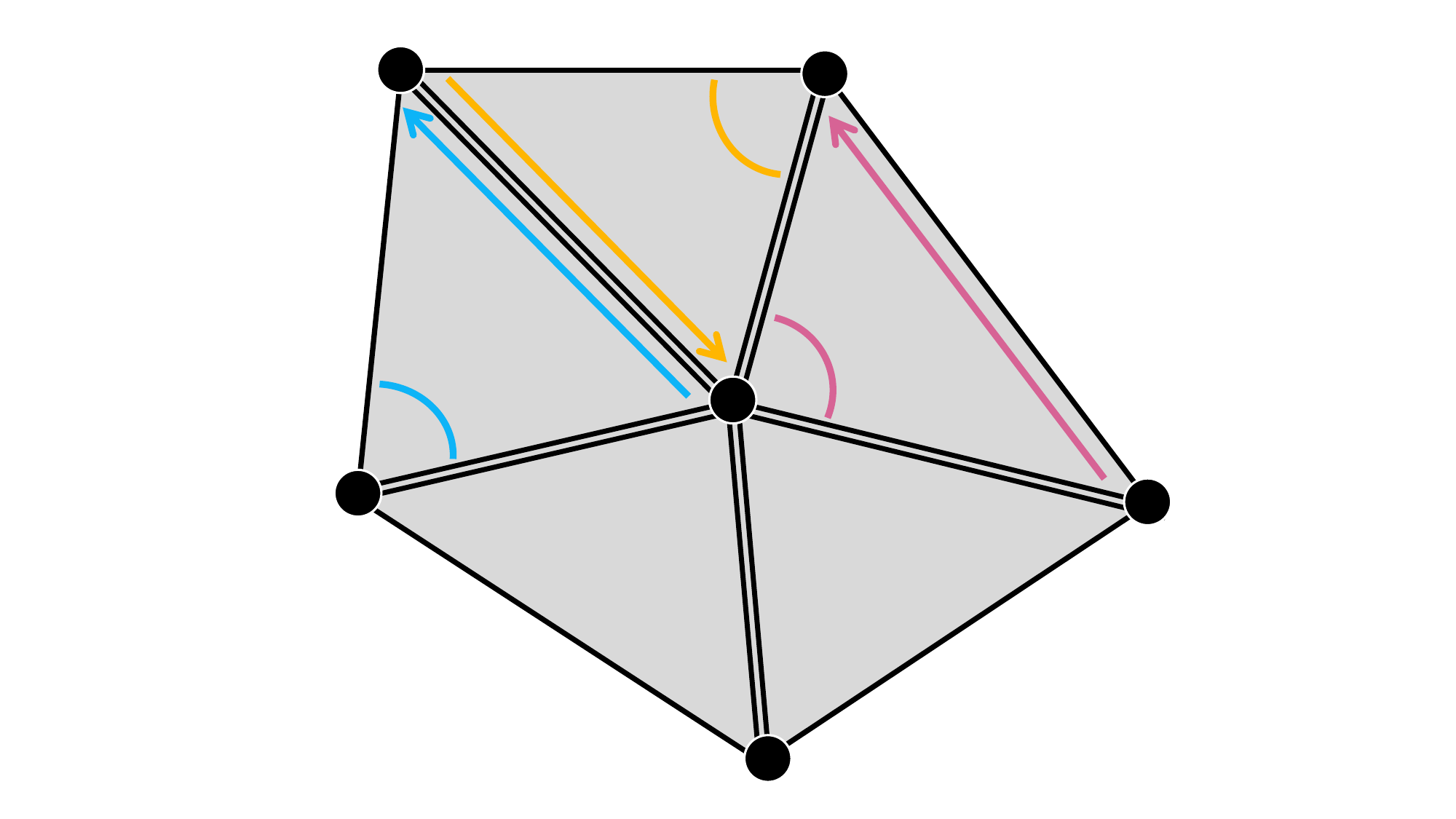}
    \put(52,23){\footnotesize $v$}
    \put(41.5,41.5){\footnotesize $\mathbf{e}$}
    \put(49.8,48){\footnotesize $\alpha_e$}
    \put(62,3){\footnotesize $\N_v$}
    \end{overpic}
    \caption{Illustration of the spokes and rims neighborhood $\N_v$ around some vertex $v$. The spokes are the edges that directly involve $v$ and thus appear twice in the set, represented through two different directions (or half-edges, see the blue and orange arrows). The rims, on the other hand, only appear in the set once as half-edges (one example is marked in pink). The angle opposite a half-edge used in the cotangent weight definition is marked in the same color as the half-edge. For example, the cotangent weight of the yellow half-edge $\mathbf{e}$ is $\alpha_e$, thus $w_e=\cot \alpha_e$.\\}
    \label{fig:sr_neighborhood}
\end{figure}
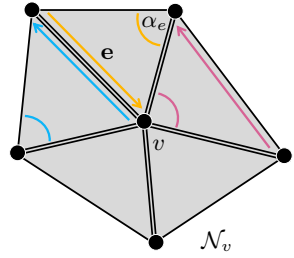

The standard ARAP energy can be defined over different neighborhoods \cite{Jacobson:FAST:2012}, such as spokes-only \cite{arap}, per-triangle  \cite{LocalGlobal:2008} or spokes-and-rims \cite{chao_sr}. We use spokes-and-rims and define the corresponding half-edge set for vertex $v$ as 
$$\N_v = \{e \in \E \ |\ e \in t \text{ and } t \in \mathcal \T_v \}.$$
The weight $w_e$ of each half-edge $e$ is 
the cotangent of the angle $\alpha_e$ opposite of $e$ in $e$'s triangle, see \figref{fig:sr_neighborhood}: 
$$w_e = \cot \alpha_e. $$
The area-corrected cotan Laplacian vector at vertex $v$ is defined in the standard way as
\begin{equation}\label{eq:lap_vec}
\boldsymbol\ell_v=\sum_{\{e \in \E|v\in e\}}\frac{w_{e}}{2 A_v}d_e^v\mathbf{e},
\end{equation}
with $d_e^v$ being the sign factor coming from the derivative of the edge vector $\mathbf{e}$ w.r.t.\ vertex $\mathbf{v}$:
\[
d_e^v =
\begin{cases}
    1 & \text{if }\mathbf{e}=(\mathbf{v}-\mathbf{u}), u \in \Vset \\
    -1 & \text{if }\mathbf{e}=(\mathbf{u}-\mathbf{v}), u \in \Vset \\
    $0$ & \text{else}
\end{cases}
\]
These weights are accumulated in the cotan Laplacian matrix $\Lap \in \Reals^{n\times n}$, such that $(\Mass^{-1}\Lap\V)_v=\lapvec_v\tran$, where $(\mathbf{X})_v$ refers to row $v$ of matrix $\mathbf{X}$. 

\paragraph{Energy and method.}
With this notation, we now define the energy we propose to minimize in order to obtain natural-looking as well as smooth results. Our smooth ARAP energy $E$ consists of two parts that are weighted against each other by the parameter $\lambda \in [0,1)$, which controls how much the smoothness of the original mesh is preserved:
\begin{equation}\label{eq:total_energy}
    E=(1-\lambda) E_\mathrm{ARAP}+\lambda E_\mathrm{smooth}.
\end{equation}
The first term corresponds to the standard ARAP energy: 
\begin{align}
    E_\mathrm{ARAP}=\sum_{v \in \Vset} A_v \sum_{e\in \N(v)} \frac{w_e}{3 A_v}  \norm{\mathbf{e}'-\smolrot_v \mathbf{e}}^2 \label{eq:arap_energy},\\
    \text{where }\smolrot_v=\argmin_{\smolrot_v \in SO(3)} \sum_{e\in \N(v)} w_e \|\mathbf{e}'-\smolrot_v \mathbf{e}\|^2. \label{eq:rotation}
\end{align}
Note that we added a coefficient of a third due to the overlap of neighborhoods, simply to get a nicer final equation to solve that does not have a scaled Laplacian matrix.
The novel higher-order term to encourage smoothness is defined such that non-rigid transformations of the Laplacian vectors (rather than edge vectors, as for standard ARAP) increase the energy: 
\begin{align}
     E_\mathrm{smooth}&=\sum_{v \in \Vset} A_v\|\lapvec_v'-\smolrot_v\lapvec_v\|^2.
\end{align}
Note that the rotation $\smolrot_v$ is the same as in the ARAP term in \eqref{eq:rotation}.

To compute the deformed mesh, we can keep the local-global strategy from ARAP:
\begin{enumerate}
    \item \textbf{Initialization:} initialize $\M'$ in some way. We follow Sorkine and Alexa \shortcite{arap} and use naive Laplacian editing for the non-interactive examples unless stated otherwise. In the interactive setting, the mesh from the previous frame is used as initialization.
    \item \textbf{Local step:} compute the optimal rotation $\smolrot_v$ for each vertex $v$ using \eqref{eq:rotation}, keeping the deformed vertex positions $\V'$ fixed (i.e., using the values from the previous iteration).
    \item \textbf{Global step:} solve a linear system for the new vertex positions $\V'$ while fixing the rotations $\smolrot_v$ computed in the local step.
    \item Go to step 2 until convergence.
\end{enumerate}

\subsection{Rotation fitting}
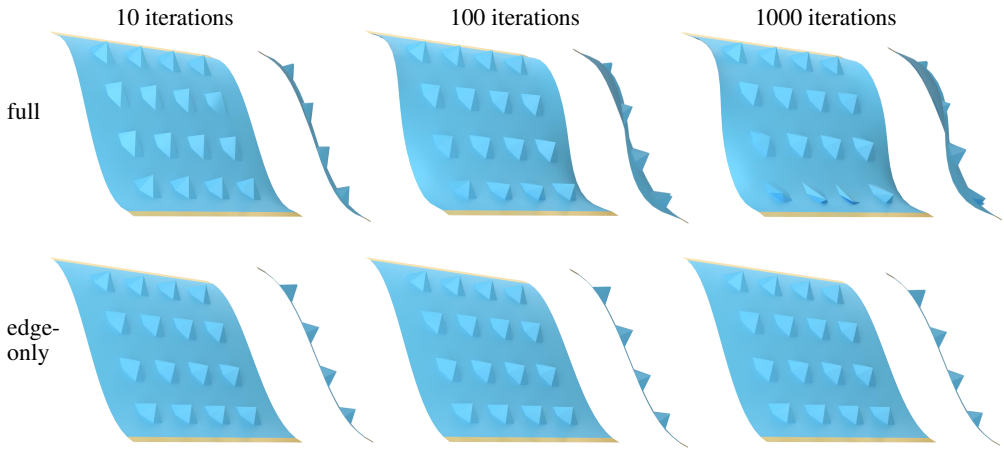
\begin{figure}[b]
\centering
\begin{overpic}[trim=0cm 0cm 0cm 0cm,width=1\linewidth,grid=false]{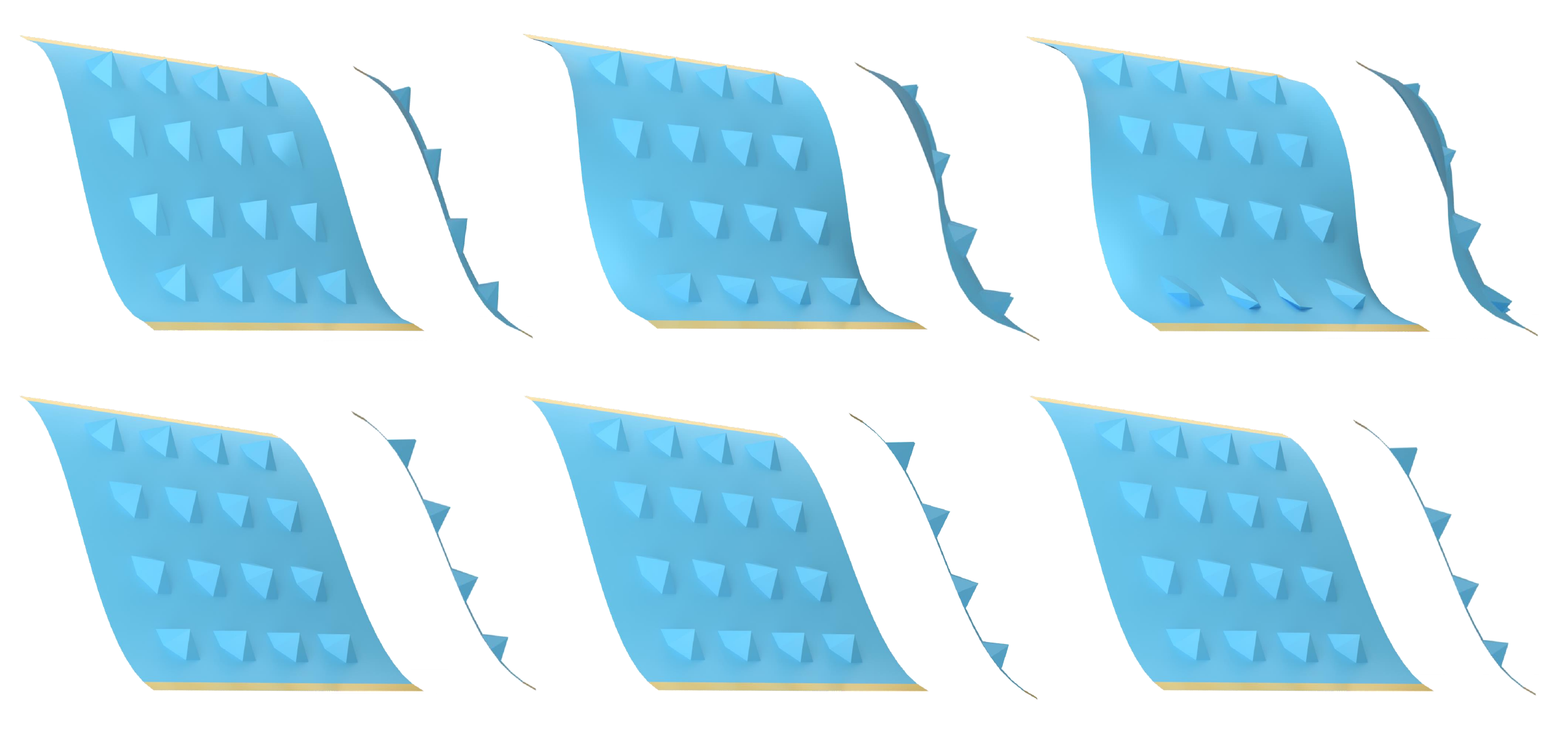}
    \put(8,45){\footnotesize 10 iterations}
    \put(42,45){\footnotesize 100 iterations}
    \put(73,45){\footnotesize 1000 iterations}
    \put(-3,35.5){\footnotesize full}
    \put(-3,13.5){\footnotesize edge-}
    \put(-3,11){\footnotesize only}
\end{overpic}
  \caption{Deformations of a plane with orthogonal spike-like details, comparing full rotation fitting (including the vertex Laplacian vector) to our proposed edge-only formulation. The results after an increasing amount of iterations are displayed from two different views, to better show the orientation of the spikes. 
  }
  \label{fig:ablation_rotation}
\end{figure}

The optimal rotation $\smolrot_v$ is determined as in the original ARAP method. Specifically, we define an auxiliary matrix $\svds$ and find the desired $\smolrot_v$ through its singular value decomposition $\svds=\svdu\svdsig\svdv\tran$.

\begin{gather*}
    \svds=\sum_{e\in \N(v)} w_e\, \mathbf{e} \, \mathbf{e'}\tran=\svdu \svdsig \svdv \tran.\\
    \smolrot_v=\svdv \begin{pmatrix} 1 & & \\ & 1 & \\ & &  \det \svdv \svdu \tran \end{pmatrix} \svdu \tran.
\end{gather*}
Please refer to the original paper \cite{arap} for more details.

\begin{figure}
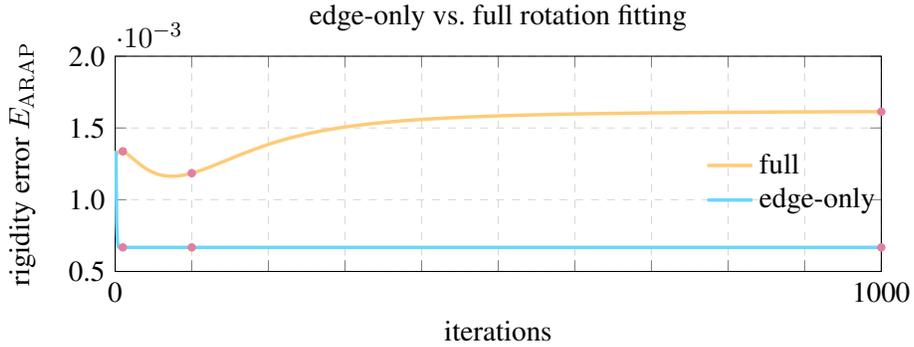

\centering
\definecolor{mycolor3}{HTML}{e080a0}
\definecolor{mycolor2}{HTML}{6cd5ff}
\definecolor{mycolor1}{HTML}{ffcb77}
\definecolor{mycolor4}{HTML}{99d98c}

\begin{tikzpicture}
    \begin{axis}[
        width=0.9\linewidth,
        height=0.34\linewidth,
        xlabel={iterations},
        ylabel={rigidity error $E_\mathrm{ARAP}$},
        legend cell align={left}, xtick={0,100,200,300,400,500,600,700,800,900,1000},
        xticklabels={0,,,,,,,,,,1000},
        ytick={0.0005,0.001,0.0015,0.002},
        yticklabels={0.5,1.0,1.5,2.0},
        xmin=0, xmax=1000,
        ymin=0.0005, ymax=0.002,
        grid=both,
        grid style={dashed, gray!30},
        legend style={at={(1.01,0.6)}, anchor=north east, draw=none, fill=none},
        title={edge-only vs.\ full rotation fitting},
    ]

    \input{figures/tikz_plot/full}    
    \input{figures/tikz_plot/edge_only};
    \addplot[color=mycolor3, only marks, mark=*, mark options={scale=0.7}] coordinates {
        (10, 0.000667655) 
        (10, 0.00133731) 
        (100, 0.000667723) 
        (1000, 0.000667723) 
        (100, 0.00118502) 
        (1000, 0.00161363)
    };
    
    \end{axis}
\end{tikzpicture}
\caption{Convergence behavior of the spiky plane experiment (cf.\ \figref{fig:ablation_rotation}) in terms of rigidity (original ARAP energy $E_\mathrm{ARAP}$) for both versions of rotation fitting. Pink dots mark the results shown in \figref{fig:ablation_rotation}.}
\label{fig:rot_energy}
\end{figure}

Note that in the definition of $\smolrot_v$ in \equref{eq:rotation}, the rotation is only fitted to the original, non-smooth term. Thus, the local step is not guaranteed to be optimal for the energy $E$ as a whole, as it might increase the value of the smooth term $E_\mathrm{smooth}$. We call the rotation fitting from ARAP \emph{edge-only}, as it considers only the mesh edge orientations for the rotation. To make the local step optimal for the whole energy, one would also have to consider the Laplacian vectors from the smooth energy term. We call this \emph{full} rotation fitting in the following. While the standard SVD approach can be extended to also include Laplacian vectors to get the result of the full rotation fitting, there are some downsides to it. 
In certain settings, visible artifacts appear. An issue of this approach is that it is always possible to perfectly align the original Laplacian to the new one, so the global optimization can only work on the scale of the Laplacians rather than their orientation when the smoothness $\lambda$ is very high, as well as it not having any direct awareness of bending in the local step. 
In practice, this can result in unintuitive rotations, as depicted in \figref{fig:ablation_rotation}. It can be seen that the full rotation fitting does not enforce rigidity well, as the spikes do not preserve their local orientation to the mesh. \figref{fig:rot_energy} shows that the rigidity is actually lower than for the initialization (even though the total energy $E$ did decrease). While a trade-off between smoothness and rigidity is expected, the observed behavior in this case is undesirable. 

Using rotations fitted only to edges does a much better job at preserving the orientation of local details. Another observation is that the mesh curves more around the border when using full rotation fitting (see \figref{fig:ablation_rotation}). Note that this curving at the border also happens with Laplacian editing \cite{LaplacianMeshEditing:2004}, see e.g.\ the bumpy plane in \figref{fig:survey_comparison}. While in some other cases, full rotation fitting does not present any observable artifacts, we never found it to be beneficial to the result, and it generally poses a disadvantage regarding the convergence and stability of the method. This can also be seen in the graph in \figref{fig:rot_energy}, where the energy when using full rotation fitting needs significantly longer to converge. Even though the edge-only rotation fitting does not use the optimal rotations for the smooth term in the local step, we have not observed energy fluctuations when evaluating the total energy with this rotation. Therefore, edge-only rotation fitting is our chosen approach.

\subsection{Vertex position optimization}
To deform the mesh, we need to find the new vertex positions $\V'$ that minimize the energy $E$. This can be achieved by differentiating the energy w.r.t.\ $\V'$. For the original ARAP term, the derivative has the following form:
\begin{equation}
    \frac{\partial}{\partial \V'} E_\mathrm{ARAP}=2\Lap\V'- 2\mathbf{b},
\end{equation}
where $\mathbf{b}$ is a collection of the original edges rotated by the rotations determined in the local step. More specifically, row $p$ of $2\mathbf{b}$ contains the derivative of $E_\mathrm{ARAP}$ w.r.t.\ vertex $p$, so
\begin{equation*}
    \mathbf{b}_p=\sum_{v \in \Vset} \sum_{e \in \N(v)}  \frac{d_e^p w_e}{3} \smolrot_v \mathbf{e}.
\end{equation*}
We rewrite the smooth energy term using a few matrices and an area-weighted Frobenius norm $\|\mathbf{B}\|_{FM}^2 := \sum_{i,j} A_i \|b_{i,j}\|^2$:
\begin{align}
     E_\mathrm{smooth}&=\sum_{v \in \Vset} A_v\|\lapvec'_v-\smolrot_v\lapvec_v\|^2 \\
        &=\|\Lap \V' - \R\|_{FM}^2, 
\end{align}
where $\R$ is constant in the wanted $\V'$, and each row $i$ corresponds to the Laplacian $\lapvec_i$ being rotated by $\smolrot_i$, so $(\R)_i=(\smolrot_i\lapvec_i)\tran$ holds. We calculate the derivative using these defined matrices:
\begin{equation}
    \frac{\partial}{\partial \V'} E_\mathrm{smooth}=2 \Lap\tran  \Mass^{-1} \Lap \V' - 2 \Lap\tran  \Mass^{-1} \R.
\end{equation}
Putting the two terms together and setting them to $0$ to find the minimum, the global step amounts to solving the following sparse linear system:
\begin{equation}
    \left (\lambda  \Lap\tran  \Mass^{-1} \Lap+(1-\lambda)\Lap\right )\V'=  \lambda   \Lap\tran  \Mass^{-1} \R+(1-\lambda)\mathbf{b}.
    \label{eq:system}
\end{equation}
{To incorporate the handle positions, we solve this system in a constrained manner using substitution. This amounts to erasing the rows and columns corresponding to constrained positions and updating the right-hand side of the equation accordingly. This is a common approach in deformation that is also taken by the original ARAP method \cite{arap}.}

\subsection{Smoothness control}\label{sec:lambda}
\begin{figure}
    \centering
    \begin{overpic}[trim=1cm 0cm 0cm 0cm,width=1.1\linewidth,grid=false]{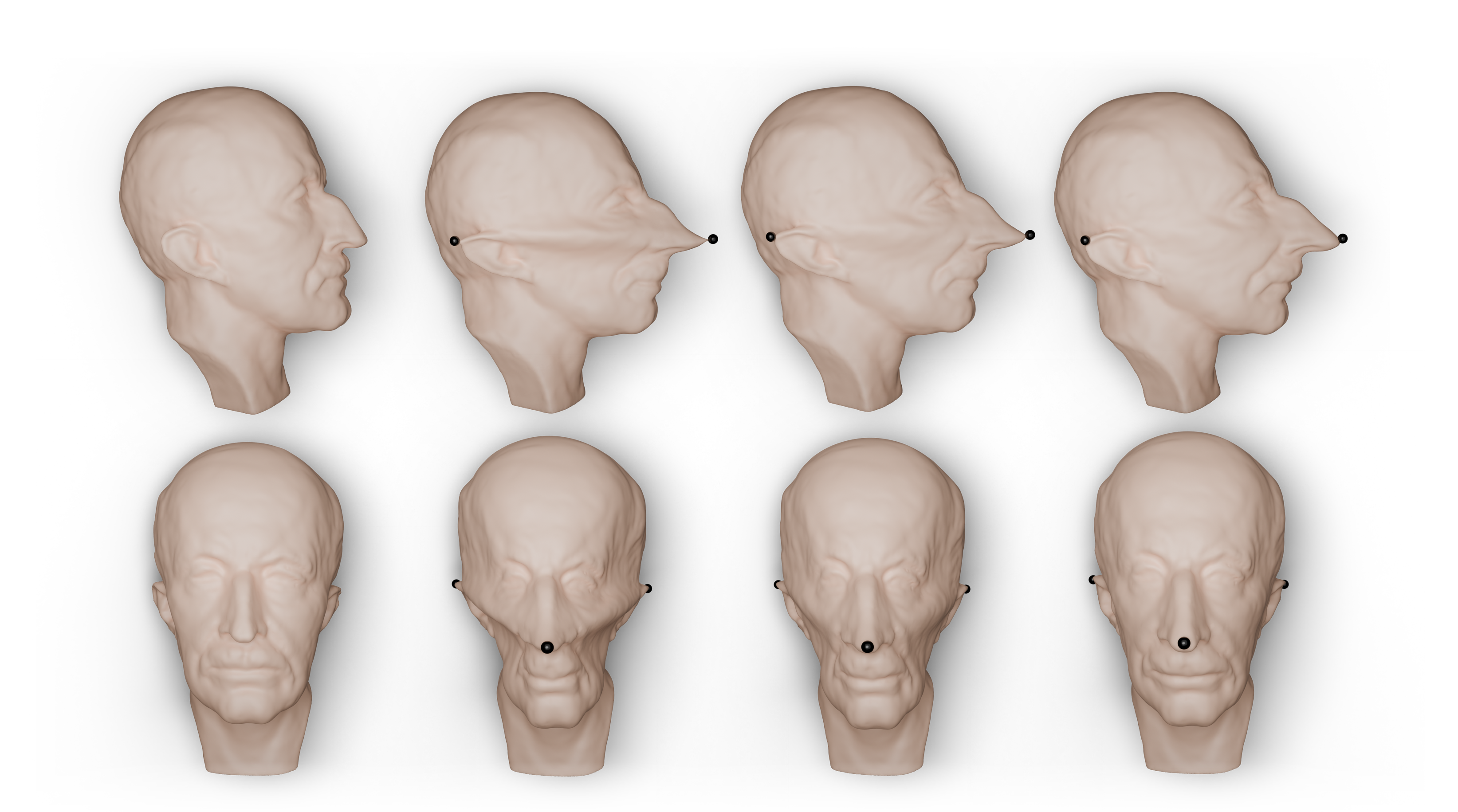}
    \put(8,55){\footnotesize original mesh}
    \put(32,55){\footnotesize $\lambda=0$}
    \put(54,55){\footnotesize $\lambda=0.5$}
    \put(75,55){\footnotesize $\lambda=0.95$}
    \put(0,38){\footnotesize side}
    \put(0,35.5){\footnotesize view}
    \put(0,16){\footnotesize front}
    \put(0,13.5){\footnotesize view}
    \end{overpic}
    \caption{Influence of smoothness parameter $\lambda$ when translating a single vertex on the nose of the Max Planck mesh away from the face, while a point on each ear is fixed, as marked by the black spheres.
      \label{fig:max_planck} }
\end{figure}
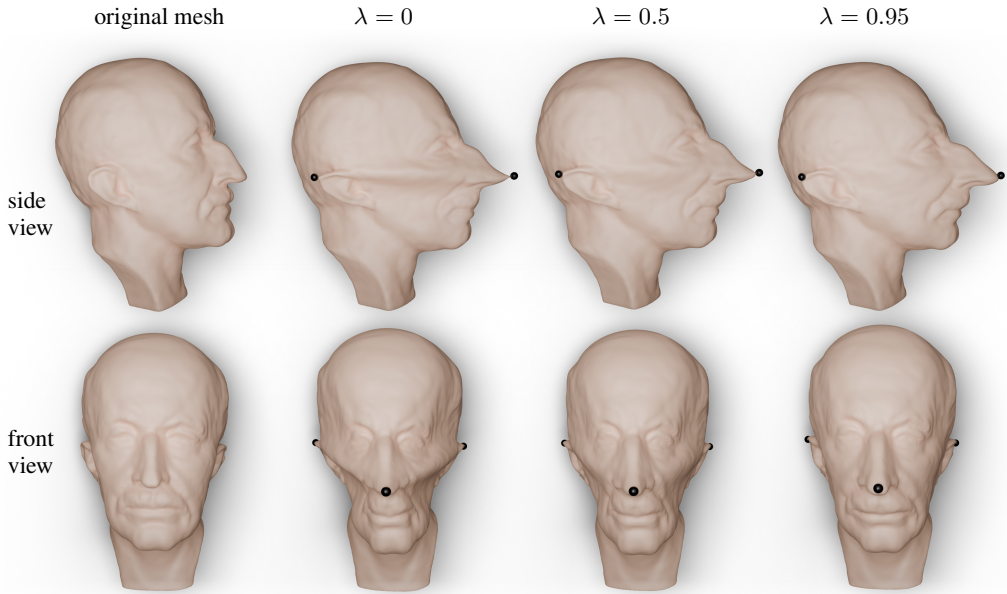

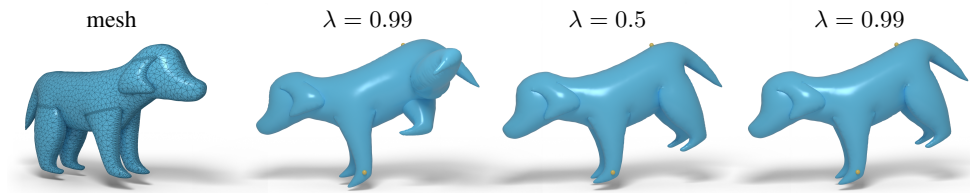
\begin{figure}
    \centering
    \begin{overpic}[trim=0cm 0cm 0cm 0cm,clip,width=1\linewidth,grid=false]{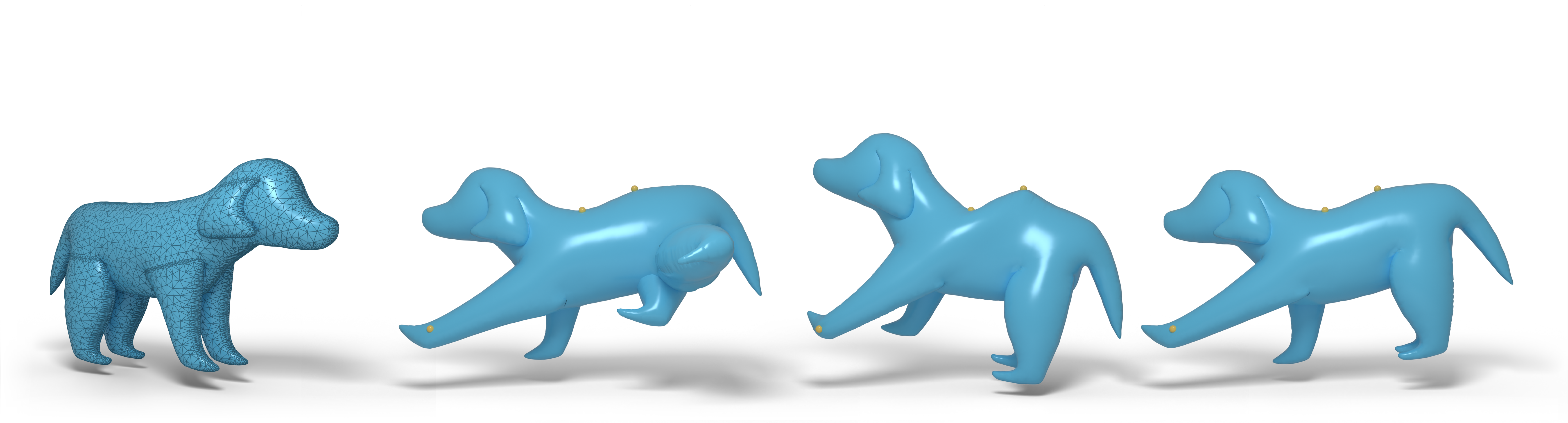}
    \put(4,23){\footnotesize a) original}
    \put(7,20){\footnotesize mesh}
    \put(24,23){\footnotesize b) interactively deformed }
    \put(27,20){\footnotesize from a) with $\lambda=0.99$}
    \put(53,23){\footnotesize c) let b) converge}
     \put(56,20){\footnotesize with $\lambda=0.5$}
    \put(75,23){\footnotesize d) let c) converge }
     \put(78,20){\footnotesize with $\lambda=0.99$}
    \end{overpic}
    \caption{Influence of smoothness parameter $\lambda$ to regularize deformations. When dynamically deforming difficult meshes such as the dog, whose highly irregular triangulation can be seen in a), attached parts can rotate more than necessary, as can be seen in b). Lower smoothness, such as $\lambda=0.5$, can recover a better solution again, see c). Initializing with that, the smoothness can also be reset and still produces nice results, as can be seen in the final picture d) with $\lambda=0.99$. }
      \label{fig:dog_regularization} 
\end{figure}

The user-chosen deformation parameter $\lambda$ controls the smoothness. In general, the higher the chosen $\lambda$, the smoother the result. This effect can be observed in \figref{fig:max_planck}, where it can be seen that increasing values of $\lambda$ both reduce the spikes around the handles, as well as limit how much the face can cave in due to the deformation.
On the other hand, $\lambda$ may also be used to regularize the deformation result. For most meshes, there are no problems when choosing an arbitrarily large $\lambda$. However, in certain challenging situations, setting $\lambda$ too high can result in unnecessary rotations. This can happen for meshes that have long parts attached via very small areas, so the rotation of the long appendages can happen without having much impact on the overall energy. One such example is the dog model from Monster Mash \cite{MonsterMash:2020}, see \figref{fig:dog_regularization}. We only observed this phenomenon in the interactive setting, where it is easier for the system to get stuck in a bad local minimum while deforming the mesh. 
In these special cases, reducing $\lambda$ helps mitigate the problem: in the mentioned example in \figref{fig:dog_regularization}, this lets the dog's legs converge to a more natural position. Having escaped the bad local minimum, the user is often free to increase $\lambda$ again if a smoother result is desired, as also shown in the figure. Thus, choosing a non-zero value for the original term is beneficial, and for challenging meshes in dynamic settings it is advised to use a $\lambda$ significantly lower than $1$. In general, no tuning is required for the smoothness parameter however, as for most meshes, choosing a high value is fine. In most examples in this paper, we opted for $\lambda=0.95$, but for increased smoothness in very high-resolution meshes, one may also increase it further. In the challenging interactive examples described before, $\lambda$ should be lowered (for example to $0.7$) if artifacts appear, but while $\lambda=0.99$ seems to create problems for some extreme deformations, we do not observe the same for $\lambda=0.95$, so one can usually still use high smoothness.

\subsection{Linear solves with efficient updates}\label{sec:solve}
Performing interactive deformations amounts to repeated solves of the linear system in \equref{eq:system}, which we abbreviate as $\mathbf A \mathbf V' = \mathbf r$ here. 
We regularize the system to remove its rank deficiency by encouraging solutions to stay close to the previous iteration:
\begin{equation} \label{eq:regularized} \tilde{\mathbf A} \mathbf V' = \tilde{\mathbf r} \quad \text{with} \quad \ \tilde{\mathbf A} = \mathbf{A}+\epsilon \mathbf{I}, \quad \tilde{\mathbf r}= \mathbf{r}+\epsilon \mathbf{V}'_\mathrm{prev}. \end{equation}
We opt for $\epsilon = 10^{-8}$, and $\mathbf{V}'_\mathrm{prev}$ denotes the vertex positions from the last iteration.
We would like to avoid refactoring the matrix each time 
constraints are added or removed. Furthermore, we aim to achieve this in a way that is solver agnostic and does not rely on special functionality for up- and downdates \cite{Davis:2009,Herholz:2020}. To this end, we consider the $n_d$ linear position constraints $\mathbf{H}  \mathbf V' = \mathbf C$, where $\mathbf C \in \Reals^{n_d \times 3}$ contains handle positions and $\mathbf H \in \Reals^{n_d \times n}$ is a sparse matrix with a single $1$ for each row selecting a constrained degree of freedom. We strive to find an approximate solution to \eqref{eq:regularized} subject to linear constraints.
Introducing a matrix $\mathcal L \in \Reals^{n_d \times 3}$ of Lagrange multipliers, we arrive at the following KKT system:
\begin{equation}
\label{eq:kkt}
\begin{pmatrix}
\tilde{\mathbf A} & \mathbf H\tran\\
\mathbf H & 0\\ 
\end{pmatrix}
\begin{pmatrix}
\mathbf V'\\
\mathcal L
\end{pmatrix} = 
\begin{pmatrix}
\tilde{\mathbf r}\\
\mathbf C
\end{pmatrix}.
\end{equation}
We define the matrix
\begin{equation}
\mathbf Q = \tilde{\mathbf A}^{-1} \mathbf H\tran \ \text{,} \quad \mathbf Q \in \Reals^{n \times n_d}
\end{equation}
and get the following result from Gaussian elimination on the KKT system \eqref{eq:kkt}:
\begin{equation}
\begin{pmatrix}
\tilde{\mathbf A} & \mathbf H\tran\\
0 & - \mathbf Q\tran \mathbf H\tran\\ 
\end{pmatrix}
\begin{pmatrix}
\tilde{\mathbf V'}\\
\mathcal L
\end{pmatrix} = 
\begin{pmatrix}
\tilde{\mathbf r}\\
\mathbf C - \mathbf Q\tran \tilde{\mathbf r}
\end{pmatrix}.
\end{equation}
The system is now solved in two steps. First, the small dense $n_d \times n_d$ system
\begin{equation}\label{eq:dense_constr} -\mathbf Q\tran \mathbf H\tran \mathcal L = \mathbf C - \mathbf Q\tran \tilde{\mathbf r}
\end{equation} 
is solved to obtain the Lagrange multipliers $\mathcal L$, which we then use to solve for $\mathbf V'$. 

This approach allows reuse of the initial factorization of $\tilde{\mathbf A}$. When constraining a new vertex, we can add a new column to $\mathbf Q$ by solving a linear system using the previously computed factorization. Removing a constraint means dropping the corresponding column in $\mathbf Q$. Note that this method enables fast updates as long as only a few dynamic constraints are present, as we have to solve the dense system in \eqref{eq:dense_constr}, which scales with the number of constraints. For the Armadillo mesh shown in Tab.\ \ref{tab:solver_timing}, the solve for the Lagrangian coefficients when having two constrained vertices takes \unit[0.868]{ms}. For fifty constrained vertices, it already takes \unit[8.26]{ms}. For the substitution approach on the other hand, the solve time goes down with the amount of constrained vertices. The presented mechanism with handle updates is thus particularly well suited for our smooth deformation energy, as it enables the use of small handles, namely point handles, without spiking artifacts.

For few constrained vertices, factorization and solving times are very similar for the substitution and updating approach. The main advantage of the updating method thus lies in the speedup of adding new handles, as can be seen in Tab.\  \ref{tab:solver_timing}. While this requires re-factorization of the system matrix for the standard approach, the updating setting avoids this and efficiently updates the constraint matrices instead.

Other solvers substitute known degrees of freedom and solve a smaller system by discarding the linear equations associated with the constrained degrees of freedom. In contrast, our approach considers the energy at constrained vertices as well.
Despite this discrepancy, we observed no visual differences in practice. Both methods are well-established for minimizing constrained energies in deformation \cite{DeformationSurvey:2008}.

\begin{table*}[t]
    \caption{Example runtimes for factorization, adding handles and solving for the usual substitution method and the one with efficient updates. Mesh complexity is given through number of vertices, runtime in milliseconds. Both meshes have two handle vertices. 
    }\label{tab:solver_timing}
    \centering
\small
{\def\arraystretch{1.0}
\setlength{\tabcolsep}{0.7em} 
\begin{tabular}{llccccccc}
\toprule[0.75pt]
mesh & \#vertices & solver & method & factorization & handle & solve\\

\midrule[0.75pt]

\multirow{4}{*}{Armadillo} 
 &  \multirow{4}{*}{172,974} & \multirow{2}{*}{Eigen} &standard&  3,358& 3,358.2 & 103.0\\
 & &  & updating & 3,341 &27.5& 104.7 \\
 \addlinespace[0.2em]
 \addlinespace[0.2em]
 & & \multirow{2}{*}{CHOLMOD} &standard&  2,043 & 2,043.4 & 78.0\\
 & &  & updating & 2,146 &44.9& 78.9 \\
 \addlinespace[0.3em]
 \cline{1-7}
 \addlinespace[0.3em]

 \multirow{4}{*}{spot} 
 &  \multirow{4}{*}{2,930} & \multirow{2}{*}{Eigen} &standard& 14& 14.0 & 1.5\\
 & &  & updating & 8 &0.8& 1.6\\
 \addlinespace[0.2em]
  \addlinespace[0.2em]
 & & \multirow{2}{*}{CHOLMOD} &standard&  7 & 7.2 & 1.4\\
 & &  & updating & 7 &1.8& 1.4 \\

 \midrule[0.2pt]

\end{tabular}
}
\end{table*}

\section{Results and discussion}
Using the presented method, we obtain deformation results that are both smooth and intuitive. In particular, local features rotate with the rest of the mesh, as is the case in the original ARAP method, and at the same time no spikes appear even with large deformations on single-point handles.

\begin{figure}[p]
    \centering
    \hspace*{-0.05\linewidth}
    \begin{overpic}[trim=0cm 2cm 0cm 0.1cm,clip,width=1.15\linewidth]{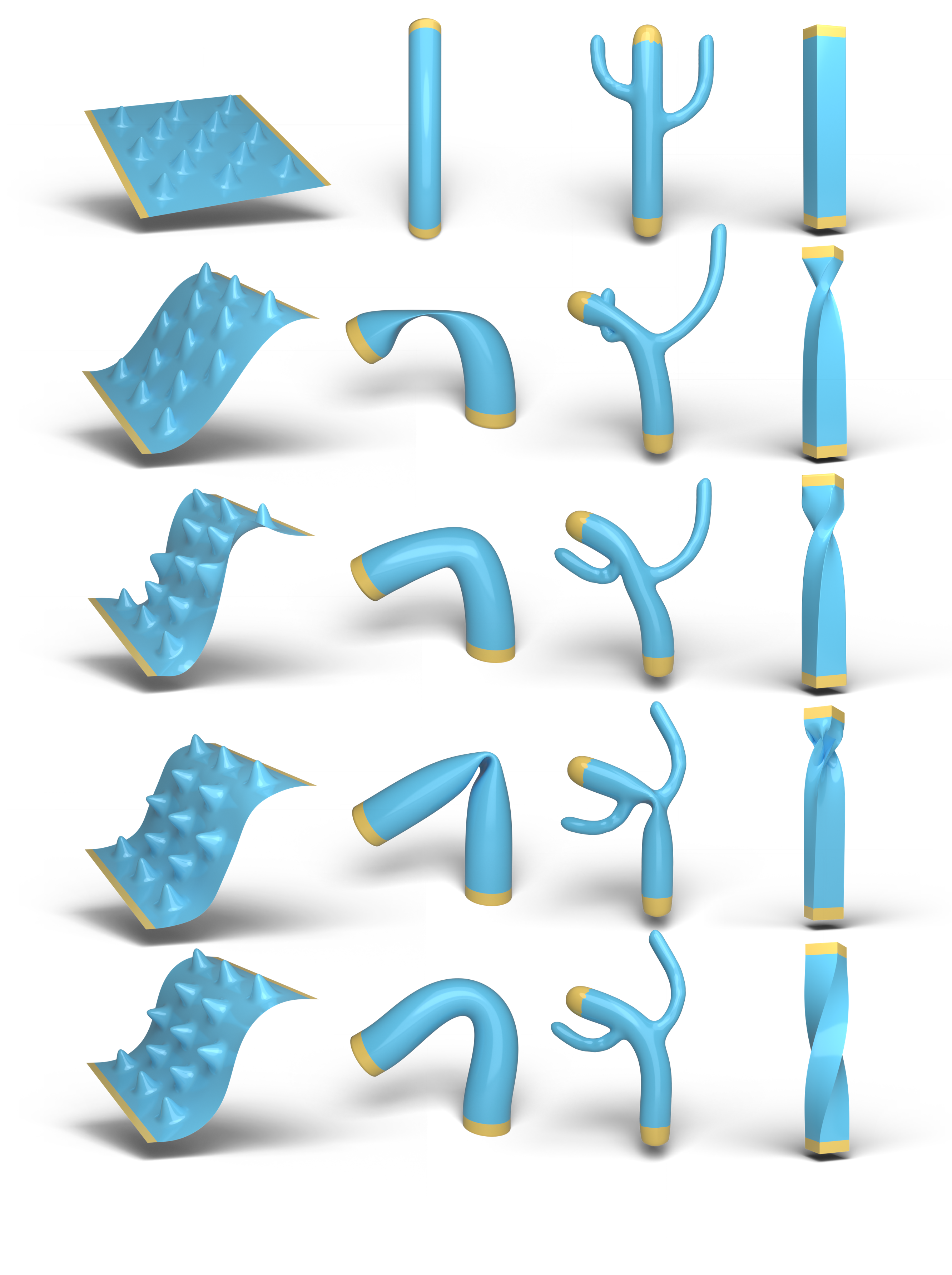}
    \put(0,88){\footnotesize Original}
    \put(0,85){\footnotesize Mesh}
    \put(0,68){\footnotesize Bi-Lap.}
    \put(0,65){\footnotesize Init.}
    \put(0,52){\footnotesize Laplacian}
    \put(0,49){\footnotesize Editing}
    \put(0,46){\footnotesize \shortcite{LaplacianMeshEditing:2004}}
    \put(0,31){\footnotesize Standard}
    \put(0,28){\footnotesize ARAP}
    \put(0,25){\footnotesize \shortcite{arap}}
    \put(0,12){\footnotesize Smooth}
    \put(0,9){\footnotesize ARAP}
    \end{overpic}
    \caption{Comparison of different deformation methods using the standard survey examples {\protect \cite{DeformationSurvey:2008}}. The handles are marked in yellow and fixed to the prescribed positions from the survey benchmark; the depicted naive Bi-Laplacian initialization is used to initialize both versions of ARAP. The results use smoothness coefficient $\lambda=0.95$ and the method is run until convergence, which we set to be a relative change of the mesh $< 10^{-4}$.}
    \label{fig:survey_comparison}
\end{figure}

A comparison of different methods on a benchmark of standard examples from the deformation survey of Botsch and Sorkine \shortcite{DeformationSurvey:2008} can be found in \figref{fig:survey_comparison}. The corresponding statistics, such as runtime and mesh size, are given in Table \ref{tab:survey}. It can be seen that our method produces very natural-looking results that don't buckle due to the increased smoothness when compared to standard ARAP. At the same time, the overall benefits of ARAP are preserved, such as the rotation of local features with the rest of the mesh.
In terms of efficiency, it can be seen that for high-resolution meshes, the smooth ARAP method needs a little longer for factorization, as the bi-Laplacian system matrix is less sparse. However, our method is usually faster overall since the smoothness regularization we use shows quicker convergence in many cases. This is the case in all examples shown in Table \ref{tab:survey}, except for the bar example, as the result of ARAP stays very close to the bi-Laplacian initialization. When using the original mesh to initialize instead, the difference becomes clearer again: the standard method takes $0.72$ seconds, while the smooth version only needs $0.45$. 

\begin{figure}[p]
    \centering
    \begin{overpic}[trim=0cm 3cm 0cm 0cm,width=1\linewidth,grid=false]{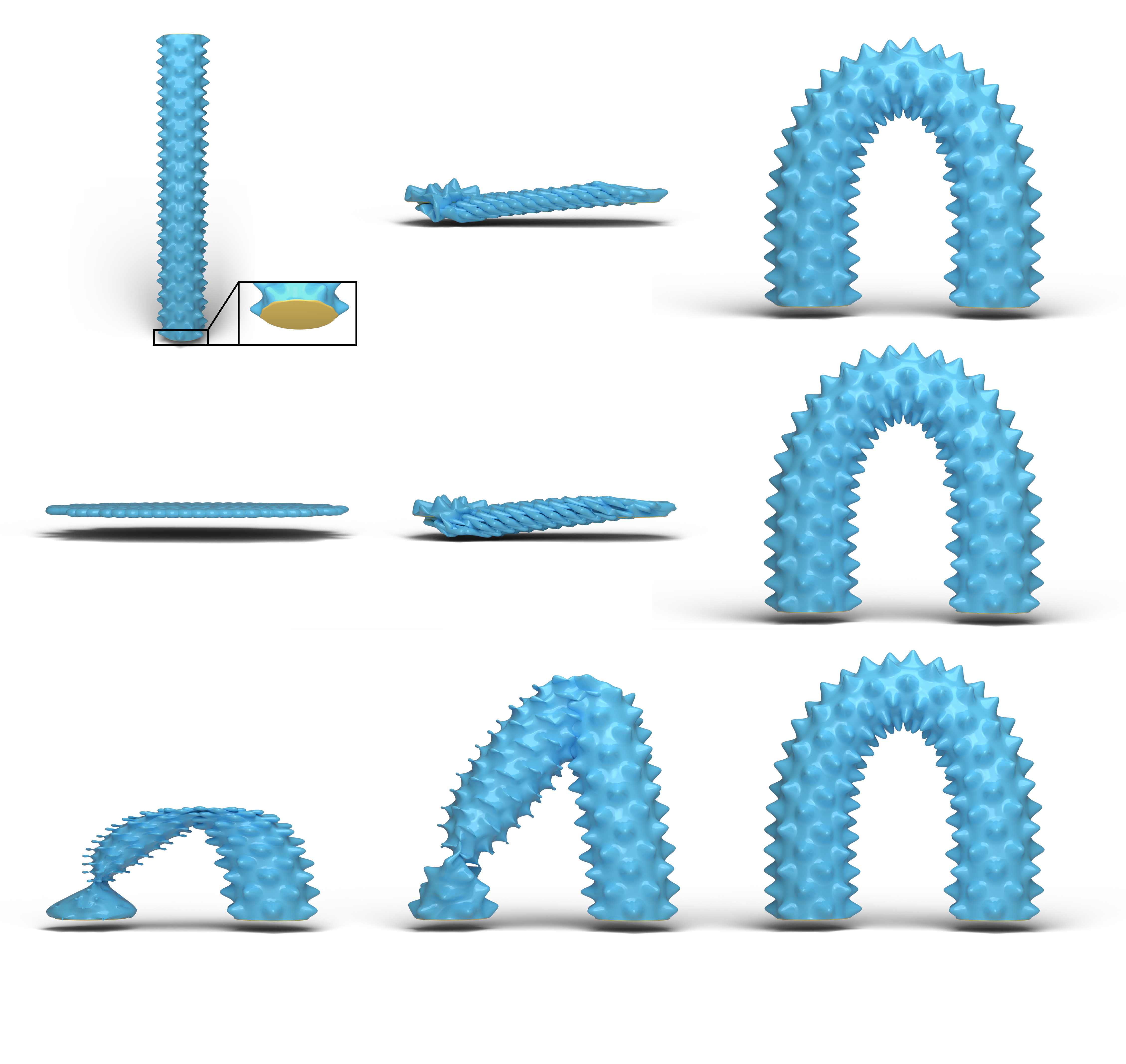}
    \put(8,86){\footnotesize Initialization}
    \put(40,87){\footnotesize Original ARAP}
    \put(36,84){\footnotesize \cite{arap}}
    \put(73,87){\footnotesize Smooth ARAP}
    \put(77,84){\footnotesize \textbf{ours}}
    \put(-8,12){\footnotesize Bi-Laplacian}
    \put(-8,41){\footnotesize Poisson}
    \put(-8,70){\footnotesize Original mesh}
    \end{overpic}
    \caption{Robustness to initialization: A cylinder with bumps is deformed through standard and smooth ARAP using different initialization methods, namely: initializing via the original mesh, solving a Poisson or a bi-Laplacian equation with constrained handle positions. Solving the Poisson equation amounts to applying Poisson mesh editing {\protect \cite{Poisson:Yu:2004}} with the positional constraints and without rotation propagation, and solving the bi-Laplacian equation is applying Laplacian mesh editing without rotation handling {\protect\cite{LaplacianMeshEditing:2004}.}
    The handles that were used are at the bottom and top of the cylinder and are again marked in yellow and are highlighted from a different perspective in the original mesh for better visibility. 
    }
      \label{fig:initialization}
\end{figure}

\begin{figure}[p]
    \centering
    \hspace*{-0.1\linewidth}
    \includegraphics[width=1.1\linewidth]{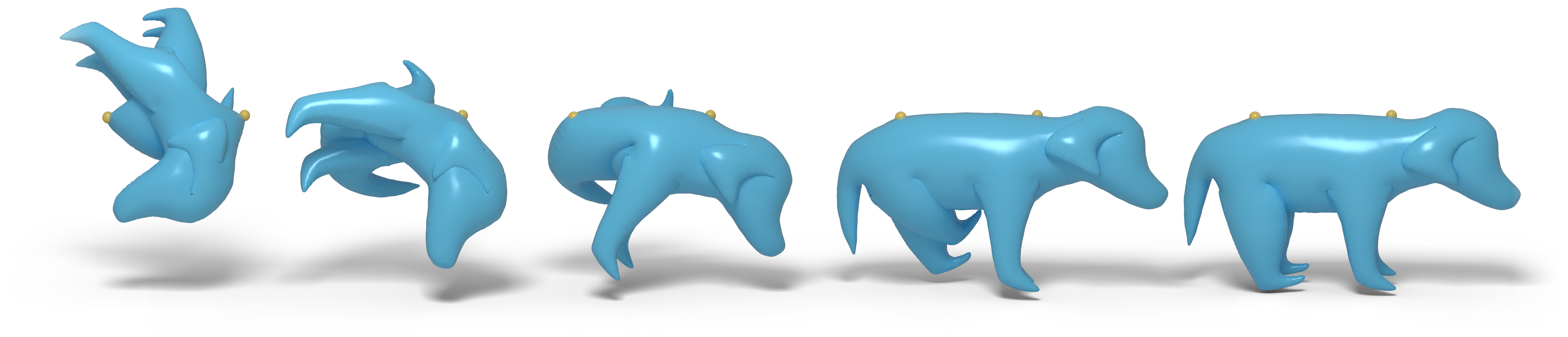}
    \caption{Robustness to bad configurations: A dog mesh is in a difficult configuration, but converges to a good result nevertheless, with some intermediate steps being shown. }
      \label{fig:robustness}
\end{figure}

Furthermore, we demonstrate the robustness of our method by showing the result of different possible initializations in \figref{fig:initialization}.  Our smooth ARAP energy is able to produce good results for very complex deformations with a significant amount of local details even from very bad initializations. This can also be seen in \figref{fig:robustness}, where an unfortunate initial configuration of the dog is shown, from which our method manages to recover by rotating the entire mesh and achieving a plausible final result that is very different to its initialization. 

Another disadvantage of our method is that it is no longer parameter-free, however, settling on $\lambda$ is very easy. For most meshes, using any high $\lambda$ such as $0.95$ works very well, and it may be further increased to for example $0.999$ should an even smoother result on a high-resolution mesh be desired. Only in very few challenging examples, $\lambda$ should not be set too high, but it can simply be reduced, should the artifacts mentioned in Section \ref{sec:lambda} be observed. 

\begin{table*}[t]
    \caption{Mesh complexity through the number of triangles and vertices, runtime (in seconds) and convergence information of the survey benchmark examples shown in \figref{fig:survey_comparison}.
    }\label{tab:survey}
    \centering
\small
{\def\arraystretch{1.0}\tabcolsep=1.0em
\begin{tabular}{llccccccc}
\toprule[0.75pt]
\multirow{2}{*}{mesh} & \multicolumn{2}{c}{complexity} & \multirow{2}{*}{method} & \multirow{2}{*}{factorization} &\multirow{2}{*}{solving}& \multirow{2}{*}{iterations}\\ \cmidrule[0.5pt](lr){2-3}
 &  \#faces & \#vtx  \\ \midrule[0.75pt]
 
\multirow{2}{*}{knubbel} 
 &  \multirow{2}{*}{80,000} & \multirow{2}{*}{40,401} &original&  0.248 & 2.44 & 61\\
 & &  & smooth & 0.253 &1.90& 44 \\
\addlinespace[0.3em]
 \multirow{2}{*}{cylinder} 
 &  \multirow{2}{*}{9,600} & \multirow{2}{*}{4,802} &original&   0.013 & 1.05&294\\
 & &  & smooth & 0.013 & 0.10 & 25 \\
\addlinespace[0.3em]
 \multirow{2}{*}{cactus} 
 &  \multirow{2}{*}{10,518} & \multirow{2}{*}{5,261} &original&  0.013 & 1.66&415\\
 & &  & smooth & 0.013 & 0.78 &  173\\
\addlinespace[0.3em]
 \multirow{2}{*}{bar} 
 &  \multirow{2}{*}{12,106} & \multirow{2}{*}{6,084} &original&   0.020 & 0.17&41\\
 & &  & smooth & 0.023 & 0.42 & 89\\
 
 \midrule[0.2pt]

\end{tabular}
}
\end{table*}

\section{Conclusion, limitations and future work}
We presented an adaptation of the classical As-Rigid-As-Possible deformation approach that produces smoother results at the constrained handles. The method benefits from the same ease of implementation as the original ARAP, with the key steps only being applying SVD in the local step and solving a sparse linear system in the global step. Our new method is robust to different initializations and is able to recover good solutions from bad starting configurations.
The added smoothness term provides more natural results in many cases and ensures that single-point control handles can be used without introducing any artifacts. This is especially beneficial in interactive settings, as point handles are the easiest interaction mechanism for user-controlled deformations.

An inherited limitation we encounter from the standard ARAP is its mesh dependence. Specifically, the neighborhoods over which rotations are fitted depend on the triangulation, so that the shapes and sizes of the 1-ring neighborhoods affect the results to varying degrees. This issue has been recently tackled via intrinsic neighborhood construction using Voronoi cells~\cite{Finnendahl:2023:AR}, and it would be an interesting avenue for future work to find a method for resolving mesh dependency without impacting the simplicity and efficiency of the deformation method.


\section*{Acknowledgements}
This work was supported in part by the ERC Consolidator Grant
No.\ 101003104 (MYCLOTH). We are grateful to Marcel Padilla for his insightful comments.


\small
\bibliographystyle{jcgt}
\bibliography{paper}

@article{Davis:2009,
author = {Davis, Timothy A. and Hager, William W.},
title = {Dynamic Supernodes in Sparse Cholesky Update/Downdate and Triangular Solves},
year = {2009},
issue_date = {February 2009},
publisher = {Association for Computing Machinery},
address = {New York, NY, USA},
volume = {35},
number = {4},
issn = {0098-3500},
url = {https://doi.org/10.1145/1462173.1462176},
doi = {10.1145/1462173.1462176},
journal = {ACM Trans. Math. Softw.},
month = feb,
articleno = {27},
numpages = {23}
}

@article{Herholz:2020,
author = {Herholz, Philipp and Sorkine-Hornung, Olga},
title = {Sparse Cholesky Updates for Interactive Mesh Parameterization},
journal = {ACM Transactions on Graphics (proceedings of SIGGRAPH Asia 2020 issue)},
volume = {39},
number = {6},
pages = {14},
year = {2020},
publisher = {ACM},
url = {https://doi.org/10.1145/3414685.3417828},
}

@inproceedings{arap,
author = {Sorkine, Olga and Alexa, Marc},
title = {As-Rigid-as-Possible Surface Modeling},
year = {2007},
isbn = {9783905673463},
publisher = {Eurographics Association},
booktitle = {Proceedings of EUROGRAPHICS/ACM SIGGRAPH Symposium on Geometry Processing},
pages = {109–116},
numpages = {8},
location = {Barcelona, Spain},
series = {SGP '07},
url = {https://dl.acm.org/doi/10.5555/1281991.1282006}
}

@article{smoothed_energy,
author = {Martinez Esturo, Janick and R\"{o}ssl, Christian and Theisel, Holger},
title = {Smoothed Quadratic Energies on Meshes},
year = {2015},
issue_date = {November 2014},
publisher = {Association for Computing Machinery},
address = {New York, NY, USA},
volume = {34},
number = {1},
issn = {0730-0301},
url = {https://doi.org/10.1145/2682627},
doi = {10.1145/2682627},
journal = {ACM Trans. Graph.},
month = {dec},
articleno = {2},
numpages = {12}
}

@Article{MonsterMash:2020,
author = "Marek Dvoro\v{z}\v{n}\'{a}k and Daniel S\'{y}kora and Cassidy Curtis and Brian Curless and Olga Sorkine-Hornung and David Salesin",
title = "{Monster Mash}: {A} Single-View Approach to Casual 3{D} Modeling and Animation",
journal = "ACM Transactions on Graphics (proceedings of SIGGRAPH ASIA)",
volume = "39",
number = "6",
articleno = "214",
year = "2020",
url = {https://doi.org/10.1145/3414685.3417805},
}

@ARTICLE{Levi_smoothed,
  author={Levi, Zohar and Gotsman, Craig},
  journal={IEEE Transactions on Visualization and Computer Graphics}, 
  title={Smooth Rotation Enhanced As-Rigid-As-Possible Mesh Animation}, 
  year={2015},
  volume={21},
  number={2},
  pages={264-277},
  url={https://doi.org/10.1109/TVCG.2014.2359463}}

@article{chao_sr,
author = {Chao, Isaac and Pinkall, Ulrich and Sanan, Patrick and Schr\"{o}der, Peter},
title = {A simple geometric model for elastic deformations},
year = {2010},
issue_date = {July 2010},
publisher = {Association for Computing Machinery},
address = {New York, NY, USA},
volume = {29},
number = {4},
issn = {0730-0301},
url = {https://doi.org/10.1145/1778765.1778775},
doi = {10.1145/1778765.1778775},
journal = {ACM Trans. Graph.},
month = jul,
articleno = {38},
numpages = {6}
}

@article{DeformationSurvey:2008,
author = {Mario Botsch and Olga Sorkine},
title = {On linear variational surface deformation methods},
journal = {IEEE Transactions on Visualization and Computer Graphics},
year = {2008},
volume = {14},
number = {1},
pages = {213--230},
url = {https://doi.org/10.1109/TVCG.2007.1054},
}

@article{LocalGlobal:2008,
  title={A local/global approach to mesh parameterization},
  author={Liu, Ligang and Zhang, Lei and Xu, Yin and Gotsman, Craig and Gortler, Steven J},
  journal={Computer graphics forum},
  volume={27},
  number={5},
  pages={1495--1504},
  year={2008},
  url={https://dl.acm.org/doi/10.5555/1731309.1731336},
}

@article{Jacobson:FAST:2012,
author = {Alec Jacobson and Ilya Baran and Ladislav Kavan and Jovan Popovi{\'{c}} and Olga Sorkine},
title = {Fast Automatic Skinning Transformations},
journal = {ACM Transactions on Graphics (proceedings of ACM SIGGRAPH)},
volume = {31},
number = {4},
year = {2012},
pages = {77:1--77:10},
url = {https://doi.org/10.1145/2185520.2185573},
}

@inproceedings{LaplacianMeshEditing:2004,
author = {Olga Sorkine and Daniel Cohen-Or and Yaron Lipman and Marc Alexa and Christian R\"{o}ssl and Hans-Peter Seidel},
title = {Laplacian Surface Editing},
booktitle = {Proceedings of the EUROGRAPHICS/ACM SIGGRAPH Symposium on Geometry Processing},
year = {2004},
pages = {179--188},
publisher = {ACM Press},
location = {Nice, France},
url = {https://doi.org/10.1145/1057432.1057456}
}

@article{Finnendahl:2023:AR,
author = {Finnendahl, Ugo and Schwartz, Matthias and Alexa, Marc},
title = {{ARAP} Revisited: Discretizing the Elastic Energy using Intrinsic {V}oronoi Cells},
journal = {Computer Graphics Forum},
year={2023},
volume = {43},
number = {5},
url = {https://doi.org/10.1111/cgf.14790},
}

@inproceedings{DeformationTutorial:2009,
author = {Olga Sorkine and Mario Botsch},
title = {Tutorial: Interactive Shape Modeling and Deformation},
booktitle = {EUROGRAPHICS},
year = {2009},
}

@article{Poisson:Yu:2004,
author = {Yu, Yizhou and Zhou, Kun and Xu, Dong and Shi, Xiaohan and Bao, Hujun and Guo, Baining and Shum, Heung-Yeung},
title = {Mesh editing with poisson-based gradient field manipulation},
year = {2004},
issue_date = {August 2004},
publisher = {Association for Computing Machinery},
address = {New York, NY, USA},
volume = {23},
number = {3},
issn = {0730-0301},
url = {https://doi.org/10.1145/1015706.1015774},
doi = {10.1145/1015706.1015774},
journal = {ACM Trans.\ Graph.},
month = aug,
pages = {644–651},
numpages = {8}
}

\section*{Index of Supplemental Materials}
The code can be found under \url{https://github.com/oehria/smooth-arap}. It contains a non-interactive implementation to recreate the survey examples with our method (see \figref{fig:survey_comparison}), the interactive framework, and the code for the solver with efficient updates. Additionally, a \href{https://youtu.be/xZV4suqSRMA?si=nzCX3intqaeeqN92}{video} is provided which showcases the interactive application in use.
\clearpage

\section*{Author Contact Information}

\hspace{-2mm}\begin{tabular}{p{0.33\textwidth}p{0.33\textwidth}p{0.33\textwidth}}
Annika Oehri \newline
Dept.\ of Computer Science \newline
ETH Zurich \newline
Switzerland \newline
\href{https://oehria.github.io/}{https://oehria.github.io/}
&

Philipp Herholz \newline
\href{https://phherholz.github.io/}{https://phherholz.github.io/}

&
Olga Sorkine-Hornung \newline
Dept.\ of Computer Science \newline
ETH Zurich \newline
Switzerland \newline
\href{https://igl.ethz.ch/people/sorkine/}{https://igl.ethz.ch/people/sorkine/}

\end{tabular}

\afterdoc

\end{document}